\documentclass[twocolumn,aps,prc,superscriptaddress,showpacs]{revtex4-1}

\usepackage{verbatim}
\usepackage{graphicx}
\usepackage{epsfig}
\usepackage{amsfonts}
\usepackage{amsmath}
\usepackage{amssymb}
\usepackage{bm}
\usepackage{hyperref}
\usepackage{color}
\usepackage{float}
\usepackage{siunitx}
\usepackage{multirow}
\usepackage{pgf}
\usepackage{soul} 
\usepackage[makeroom]{cancel} 

\DeclareUnicodeCharacter{2212}{-}

\begin{document}

\title{Neutron-pair structure in the continuum spectrum of $^{26}$O}

\author{S. Affranchino}
\affiliation{Department of Physics FCEIA (UNR),
             Av. Pellegrini 250, S2000BTP Rosario, Argentina}
\affiliation{Institute of Nuclear Studies and Ionizing Radiations (UNR), 
		    	Riobamba y Berutti, S2000EKA Rosario, Argentina.}             
		    	
\author{R.M. Id Betan}
\affiliation{Department of Physics FCEIA (UNR),
             Av. Pellegrini 250, S2000BTP Rosario, Argentina}
\affiliation{Institute of Nuclear Studies and Ionizing Radiations (UNR), 
		    	Riobamba y Berutti, S2000EKA Rosario, Argentina.}             
\affiliation{Physics Institute of Rosario (CONICET-UNR), 
             Esmeralda y Ocampo, S2000EZP Rosario, Argentina}


\begin{abstract}
  \begin{description} 
    \item[Background] The structure of $^{26}$O is currently being investigated on both theoretical and experimental fronts. It is well established that it is unbound and the resonance parameters are fairly well-known. The theoretical analysis may involved two- and three-body interactions, as well as correlations with the continuum spectrum of energy.
    \item[Purpose] In order to properly assess the structure of the ground and excited states, it is imperative to include a large single particle representation with the right asymptotic behavior. The purpose of this work is to provide details of the single particle continuum configurations of the ground and excited $0^+$ states.
    \item[Method] We use a large complex energy single particle basis, formed by resonances and complex energy scattering states, the so called Berggren basis, and a separable interaction, which is convenient to solve in a large model space.
    \item[Results] Three $0^+$ states were found in the complex energy plane. Changes of the resonant parameters, i.e. energy and width, were analyzed as a function of strength of the residual interaction. It is shown how a subtle difference in the interaction could change the unbound character of $^{26}$O into a Borromean nucleus. 
     \item[Conclusions] Only one of the two excited states can be considered as a candidate for a physical meaningful resonance. The calculated occupation probabilities are in agreement with other theoretical approaches although the calculated half live is three-order of magnitude smaller than the experimental one.
  \end{description}
\end{abstract}

%
                  
\maketitle

\section{Introduction}
The experimental discovery of the radioactive decay of the nucleus $^{45}$Fe \cite{2002Oliveira} triggered the study of the physics of two-proton decay almost two decades ago. Similarly, $^{26}$O \cite{2012Lunderberg} ignited the study of the exotic two-neutron radioactive decay, and attracted much attention in the last few years from both, theoretical and experimental side \cite{2013Caesar,2013Grigorenko,2013Kohley,2016Kondo}.

Nowadays, it is accepted that the drip line of the Oxygen chain occurs at $^{24}$O \cite{1970Artukh} since the nuclei $^{25}$O \cite{1985Langevin} and $^{26}$O \cite{1990Mueller,2013Grigorenko,2013Caesar} are unbound. Experimentally, it has been established that the energy and half-life of $^{26}$O are $18 \pm 4$ keV \cite{2016Kondo} and $4.5 \pm 3$ ps \cite{2013Kohley}, respectively. Even if better statistics would be desired, these resonant parameters can be considered reliable. An excited $2^+$ state is known to lie at the energy $1280^{+110}_{-80}$ keV \cite{2016Kondo}, which places the first excited state much closer to the continuum threshold than the previously known,  $4225^{+227}_{-176}$ keV \cite{2013Caesar}. Even where there is no experimental evidence for $0^+$ excited states, there can be theoretical predictions for them \cite{2016Hagino,2015Grigorenko}.

Many theoretical calculations have been performed regarding the structure of $^{26}$O. Some of them predicted it to be bound, for example, using Gogny \cite{2008Schunck} and semi-realistic \cite{2008Nakada}  interactions it was found that $S_{2n}$($^{26}$O)$>0$, in particular, Ref. \cite{2008Nakada} obtained $S_{2n}$($^{26}$O) $\sim 1.5$ MeV. Shell model calculations using the phenomenological USD interaction \cite{1988Brown}, also predicted it to be bound by approximately $1$ MeV but it is unbound using the USDA/B parametrization \cite{2006Brown}. In light of these results and the experimental evidence, some refinements were introduced in the theories, including three body forces and continuum coupling. For example, in Ref. \cite{2010Otsuka}, it was shown that a repulsive three-body force reconciles theory with experiment regarding the location of the drip line in the Oxygen isotopes. The new version of the Continuum Shell Model \cite{2005Volya,2006Volya}, predicted the ground state of $^{26}$O to be unbound by 0.021 MeV and its first exited state $2^+$ at 1.870 MeV, close to the experimental observations \cite{2016Kondo}. 

Other theoretical approaches were also used to describe this nucleus. Green function formalism \cite{2014aHagino,2014bHagino,2016Hagino}, continuum-coupled shell model in a spherical well \cite{2015Tsukiyama}, three-body model in the hyperspherical harmonics formalism \cite{2015Grigorenko}, Gamow Shell Model \cite{2017Fossez}, ab-initio Gamow Shell Model \cite{2020Hu}, core plus two valence particles self-consistent model \cite{2017Hove}, and pseudostate method \cite{2020Casal}. This paper intends to address some missing ingredients in the above references, in particular, the evolution of the resonances in the complex energy plane. We calculate the half live of the ground state; the continuum-continuum contribution of each partial wave up to the $0g$-shell; and develop a criterion to distinguish between physically significant excitations $0^+$ \cite{2016Hagino,2015Grigorenko} from those that are not.

In the next section \ref{sec.formalism} we introduce the three-body model Hamiltonian and the single- and two-body complex representations. In sec. \ref{sec.3} we define the mean-field interaction from the experimental  data of the nucleus $^{25}$O, while in sec. \ref{sec.4} we study the trajectory of the $0^+$ states in the two-body complex energy plane. In sec. \ref{sec.exp} we compare our results with experiment and other models. The last section \ref{sec.conclusion} gives concluding remarks.

\section{Formalism} \label{sec.formalism}
The three-body shell model Hamiltonian, in the single particle Berggren basis \cite{1968Berggren,1996Liotta}, is used to diagonalize the $^{24}$O plus two neutrons system. A Woods-Saxon plus spin-orbit is used for the mean-field, and a  separable force \cite{1971Bes,1995Vertse} is used for the two-body residual interaction. In this section we describe the key elements which make possible to track the three-body poles in the complex energy plane.

\subsection{Single-particle complex-energy representation}
Finite well potentials have a continuous spectrum of energy. It could be the case that some of these continuum states have physical relevance, like the ones represented by the poles of $S$-matrix close to real energy axis \cite{1928Gamow,1928Condon,1939Siegert}. Berggren \cite{1968Berggren} showed how to incorporate the resonances in a basis in the style of the usual real energy  case \cite{1982Newton}. Realistic calculations using the Berggren representation were first reported in \cite{1996Liotta}. One of the key features of the Berggren basis is that it incorporates resonant states, in the same footing as bound and non-resonant continuum states and all their possible combinations in the two-particle basis. In Ref. \cite{1968Berggren} Berggren showed that a set of bound, resonant and continuum complex-energy scattering states form a representation,
\begin{equation}\label{eq:relcomp}
\delta(r-r') = \sum_{n} u_{n}(r) \, u_{n}(r') + \int_{L^+} u(r,\varepsilon) \, u(r',\varepsilon) \, d\varepsilon 
\end{equation}
where the sum runs over all bound states and the resonant states enclosed by the contour $L^{+}$ and positive real energy axis. The norm of the resonant states, are between the state and its time reversed partner, i.e. the Berggren basis is actually a bi-orthonormal basis. Equally, averages are calculated between a state and its time reversed partner, so that, probabilities become complex numbers. The physical interpretation of this has been discussed in detail in Refs. \cite{1996Berggren,1999Civitarese,2001Bianchini}. In particular for narrow resonances, which tend to be physically relevant, these probabilities become almost real quantities.

\begin{figure}[h!t]
  \includegraphics[angle=-0,width=0.23\textwidth]{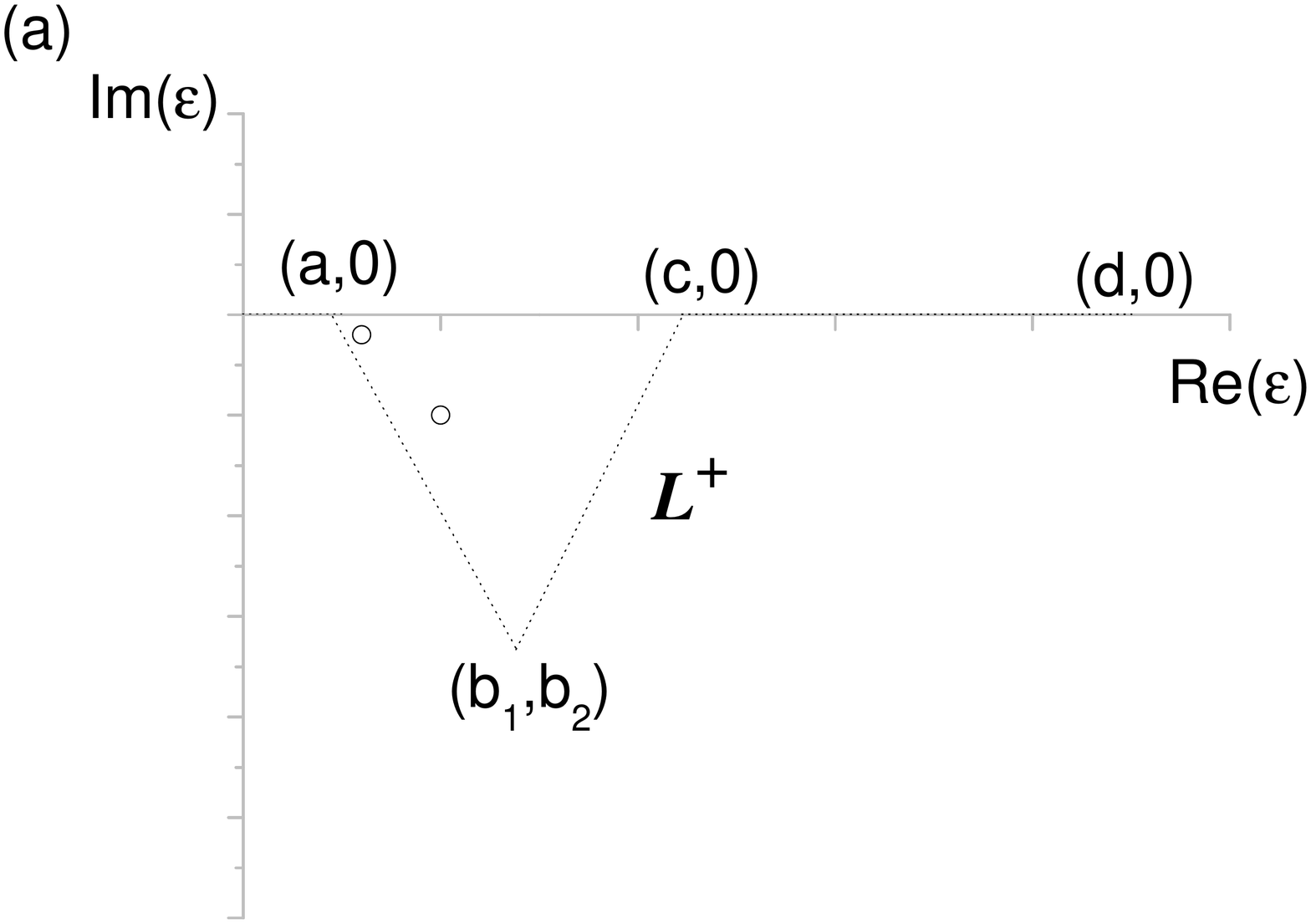}
    \includegraphics[angle=-0,width=0.23\textwidth]{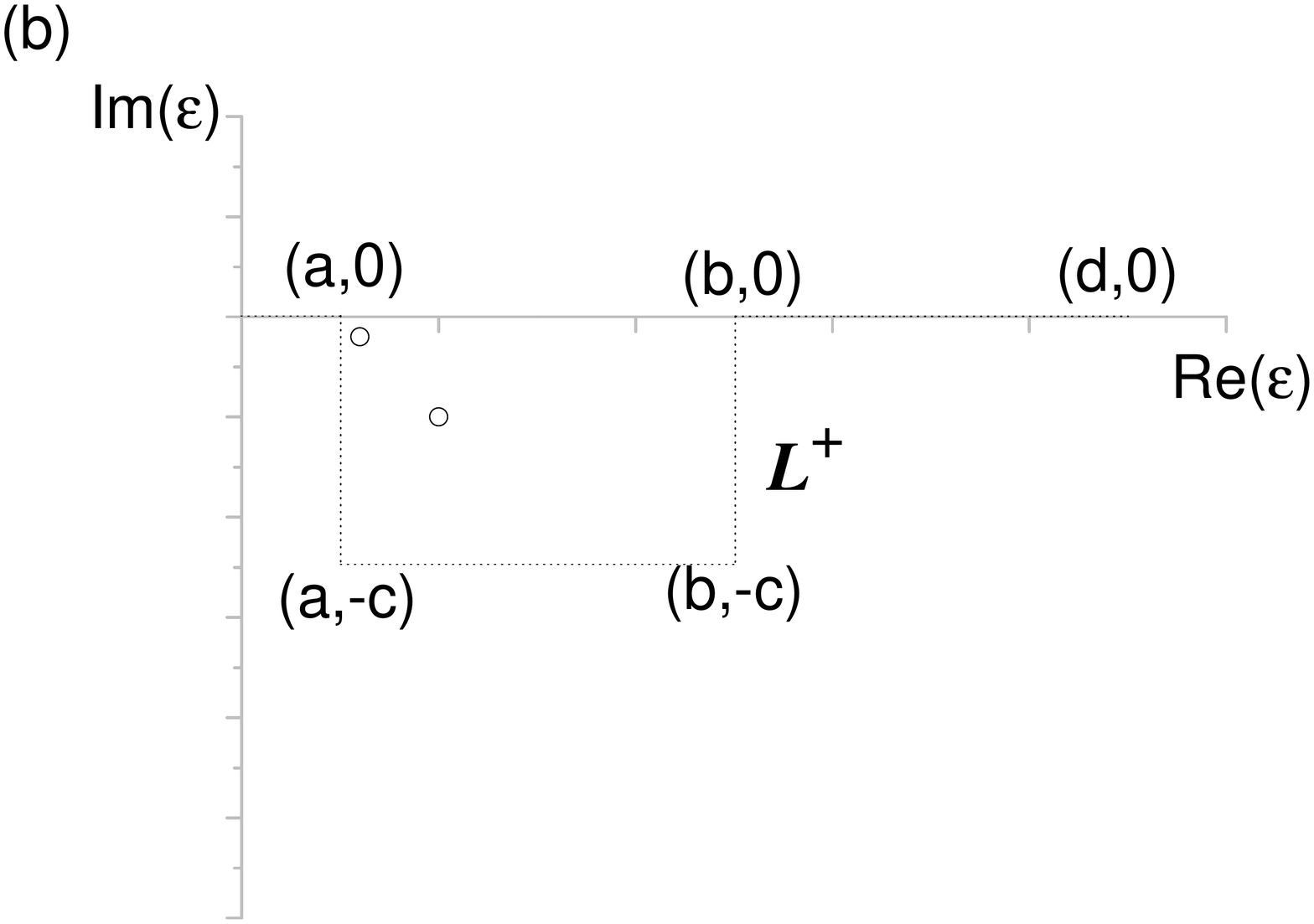} 
\caption{Examples for path $L^{+}$ to build the single-particle Berggren representation.}
\label{fig.cont1}
\end{figure}

The path $L^{+}$ is not uniquely defined \cite{1993Berggren} and different forms can be chosen according to the properties of the system under study \cite{2003IdBetan}, although it must start at the origin and finish at infinite on the real energy axis, as is shown in Fig. \ref{fig.cont1}.

For the numerical application, the integral in Eq. (\ref{eq:relcomp}) is discretized, 
\begin{equation}\label{eq:disc}
 \int_{L^+} u(r,\varepsilon) \, u(r',\varepsilon) \, d\varepsilon 
     \simeq
     \sum_{p}^{N_p} h_p u(r,\varepsilon_p) \, u(r',\varepsilon_p) 
\end{equation}
 with $\varepsilon_{p}$ and $h_{p}$ being parameters determined by the Gauss-Legendre quadrature.
 
The number of mesh-points $N_p$ for each partial wave is optimized to have the minimum number of scattering states which give stable solutions for all values of the two-body interaction strength. In this way we obtain a single-particle representation $\lbrace \vert\varphi_n\rangle ; \vert\varphi_p\rangle \rbrace$, comprised by a set of resonances, $ \langle r\vert\varphi_n\rangle = u_n(r)$ and scattering states $ \langle r\vert\varphi_p\rangle = \sqrt{h_p} \, u(r,\varepsilon_p)$.  All single-particle states are calculated using the code  ANTI \cite{1985Ixaru,1996Liotta}, which provides the continuum and pole states.

\subsection{Two-particle complex-energy representation}
In this paper we will make use of the single-particle Berggren basis to study the trajectories of the $0^+$ states of $^{26}$O and we will use a rectangular contour to have an easy way to identify the physical relevant states \cite{2002IdBetan}, i.e. to distinguish between the resonant continuum from the non-resonant continuum.

\begin{figure}[h!t]
  \includegraphics[angle=-0,width=0.23\textwidth]{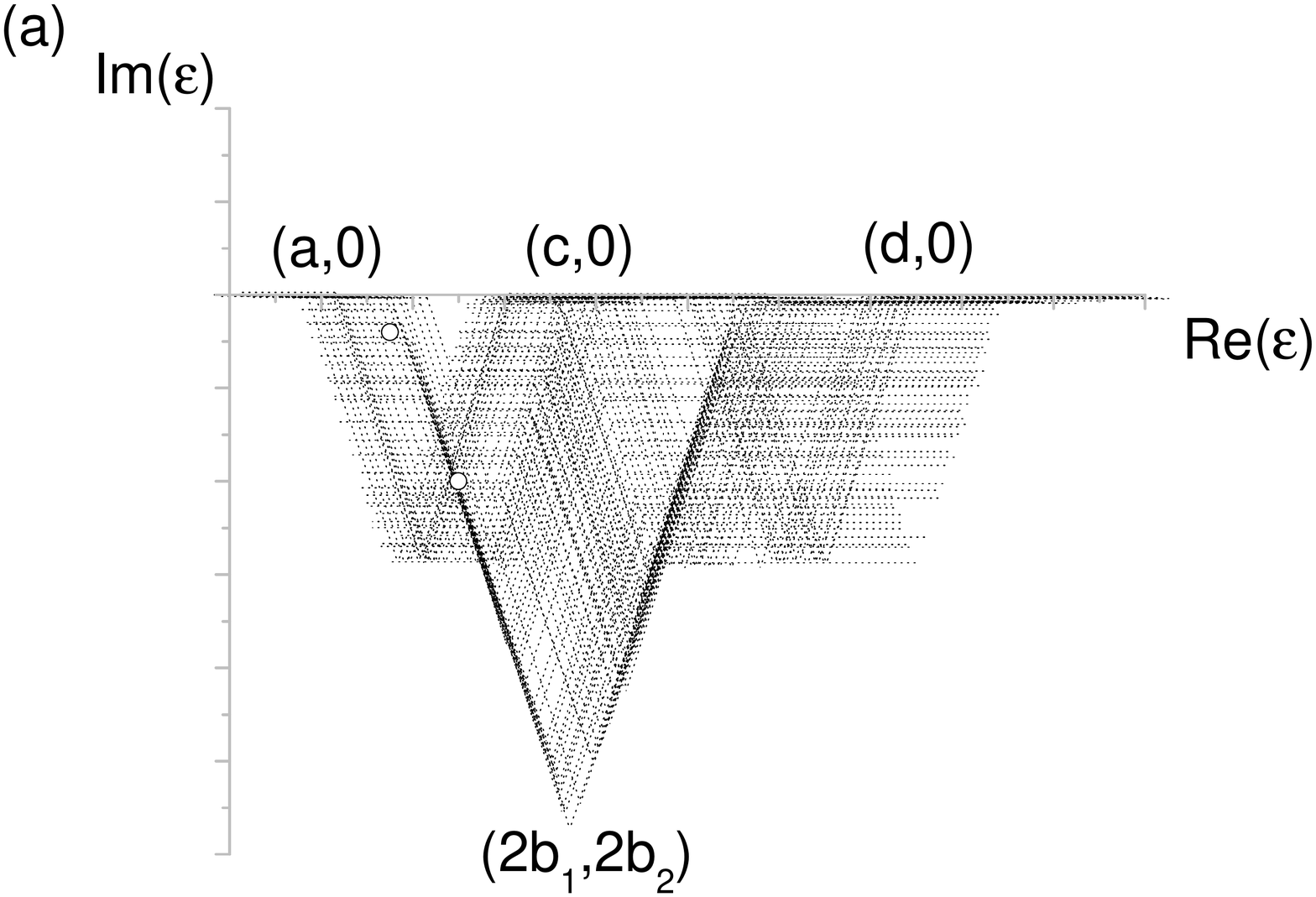}
  \includegraphics[angle=-0,width=0.23\textwidth]{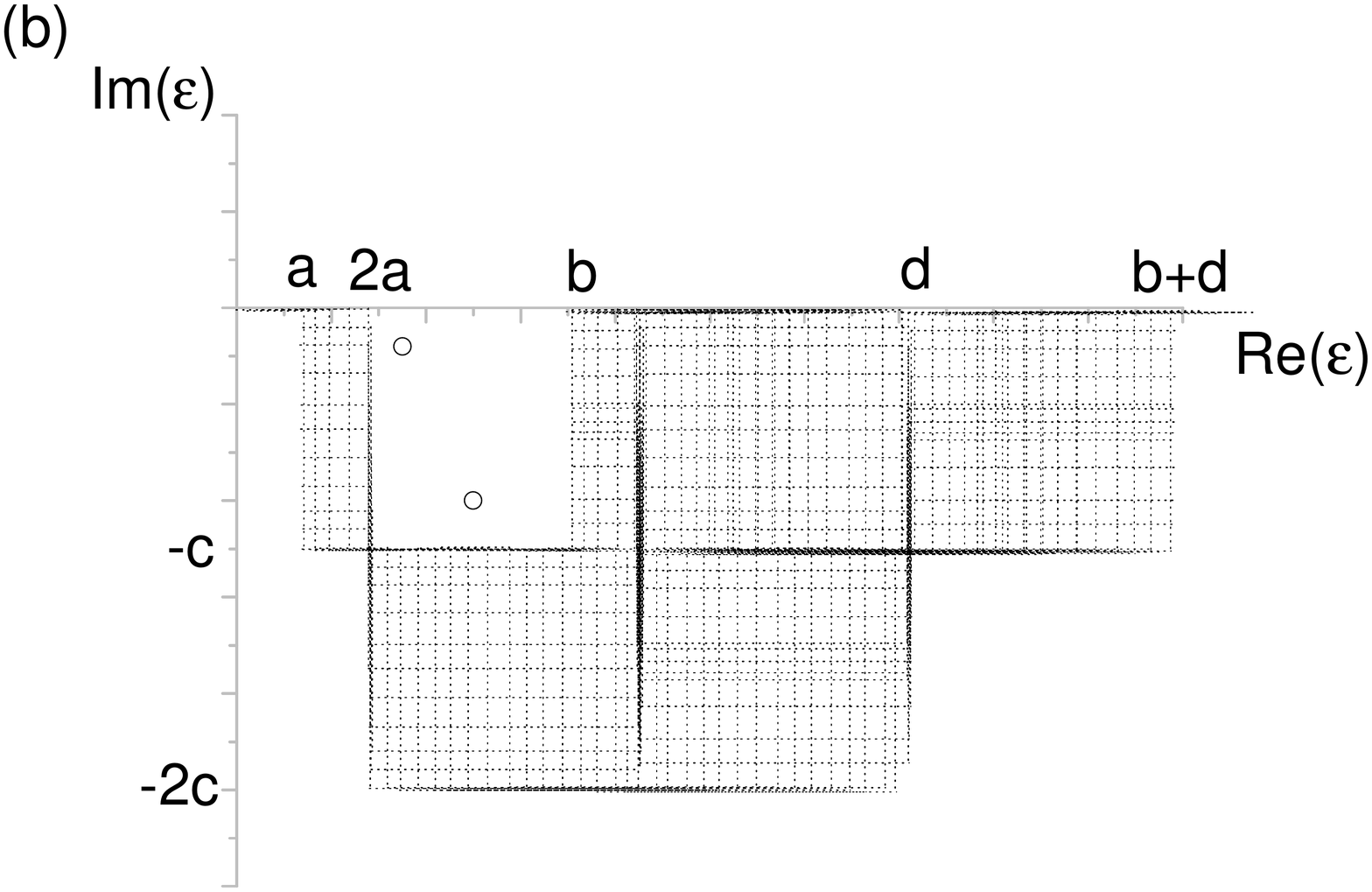} 
  \caption{Examples of two-body complex-energy representation built from the single-particle complex-energy representation of Fig. \ref{fig.cont1}.}
\label{fig.cont2}
\end{figure}

The two-particle representation is built, as usual, by taking an antisymmetric and normalized  tensor product of the above single-particle complex energy representation with itself, coupled to angular moment $J=0^+$, $| \psi^{(0)}_{ij} \rangle$ \cite{2002IdBetan,2002Michel,2009Michel}. The combination of resonant states with the contour $L^+$ of Fig. \ref{fig.cont1} produces a shifted contour in the complex energy plane, while the combination of states in the contour with itself covers a wide region, as shown in Fig. \ref{fig.cont2}. The set of zeroth-order energies $\varepsilon_i+\varepsilon_j$, with $\varepsilon_i$ and $\varepsilon_j$ both belonging to the contour $L^+$, may cover the whole complex energy plane of interest; for instance, Fig. \ref{fig.cont2}-a shows that the region of interest is full of non-resonant continuum states when the triangular contour, Fig. \ref{fig.cont1}-a, is used for $L^+$. Then, it could be difficult to find physically relevant states, with the problem becoming even more acute as the number mesh-points is increased.

From the eigenvalue equation 
$H| \Psi_\alpha \rangle = E_\alpha | \Psi_\alpha \rangle$, with the wave function 
$ | \Psi_\alpha \rangle = \sum_{i\leq j} X_{ij,\alpha} \vert\psi^{(0)}_{ij}\rangle$, 
for the $0^+$ states, we get the following secular equation to obtain the complex energy $E_\alpha$ and the  complex coefficients $X_{ij,\alpha}$, 
\begin{equation} \label{eq.shell1}
  \left( E_\alpha - \varepsilon_{i} - \varepsilon_{j} \right) X_{ij,\alpha} 
        - \sum_{k \le l}
           \langle \tilde{\psi}^{(0)}_{kl} \vert V \vert \psi^{(0)}_{ij}\rangle
           X_{kl,\alpha} = 0
\end{equation}
where $\alpha$ label the different $0^+$ states, and the tilde is to reference that the average is done between time reversed states \cite{2003IdBetan}. The label $\{k,l\}$ means $k=\{ n_k, l_k, j_k \}$ and $l=\{ n_l, l_k, j_k \}$, since only the principal quantum number $n$ is different because of the $J=0^+$ coupling.  

There is large number of correlated energies which one gets from the above equation, but only a few will be of physical interest. In order to easily identify them we choose a rectangular contour for the path $L^+$ in Eq. (\ref{eq:relcomp}), defined by the points $P_{0}=(0,0), P_{1}=(a,0), P_{2}=(a,-c), P_{3}=(b,-c), P_{4}=(b,0)$ and $P_{5}=(d,0)$ (Fig. \ref{fig.cont1}), with $d$ the single-particle energy cut-off. This path leaves a zone in the two-particle complex energy plane, between the real energy values $2a$ and $b$ and imaginary energy value $c$, almost free of zeroth order states, where the expected two particle states may lie (Fig. \ref{fig.cont2}). 

For the residual interaction we use a separable force \cite{1971Bes} since it significantly simplifies the secular equation (\ref{eq.shell1}) \cite{1995Vertse,2012prcIdBetan,2012Mukhamedzhanov}, by changing the complex-matrix diagonalization into a complex-root finding, 
\begin{equation}\label{eq.sepforce}
   \langle \tilde{\psi}^{(0)}_{kl} \vert V \vert \psi^{(0)}_{ij}\rangle = - G M_{kl} M_{ij}
\end{equation}
with 
$M_{kl} = f_{kl}  \langle k || Y_{0} || l \rangle$, $f_{kl}=\int dr\, u_k(r) f(r) u_l(r)$. The form factor $f(r)$ is
\begin{align*}
    f(r)= \left\{
    \begin{array}{ccc}
         \frac{r \partial U}{\partial r}  & \hspace{3mm} & \textnormal{Surface interaction} \\
         U(r) & \hspace{3mm} & \textnormal{Volume interaction}
    \end{array}
    \right.
\end{align*}
where $U(r)=(1+exp{\frac{r-R}{a}})^{-1}$. The reduced matrix elements are, 
$\langle k || Y_{0} || l \rangle = \frac{ (-)^{ j_k+1/2} (2j_k+1)}{ \sqrt{4\pi}} \langle j_k \frac{1}{2} 0 0 | j_k, -\frac{1}{2} \rangle$ \cite{2003IdBetan}.

From Eqs. (\ref{eq.shell1}) and (\ref{eq.sepforce}) we obtain the dispersion relation which will be used to evaluate the correlated energies $E_{\alpha}$. Because we are using the Berggren metric, the square of the matrix element appears instead of the square of the absolute value.
\begin{equation}\label{eq.dispersion}
 \dfrac{1}{G} = - \sum_{k \leq l} \dfrac{M^2_{kl}}{E_{\alpha} - \varepsilon_k - \varepsilon_l} 
\end{equation}

The strength $G$ is varied from zero to some maximum value with the aim of following the evolution of each one of the $0^+$ state of the model space. For each energy $E_\alpha$ we calculated the wave function amplitude with the equations
\begin{align} \label{eq.x}
   X_{ij,\alpha} &= N_{\alpha}  \dfrac{M_{ij}}{E_{\alpha} - \varepsilon_i - \varepsilon_j} 
\end{align}
where $N_{\alpha} $ is the normalization constant such that $\sum_{i \le j }X^2_{ij,\alpha}=1$.

\section{Results} \label{sec.results}

\subsection{Mean-field and complex single-particle basis} \label{sec.3}
Since the nucleus $^{25}$O is unbound there is no bound state in the single-particle basis, i.e. it only contains continuum states. The mean-field parameters are fitted using $\chi^2$ optimization in order to reproduce the resonant parameters of the ground state $3/2^+$ of $^{25}$O and the gap with the hole state $1/2^+$. From Ref. \cite{2008Hoffman} we have $\varepsilon_{1s_{1/2}}=-4.09 \pm 0.13$ MeV, and from Ref. \cite{2016Kondo} $\varepsilon_{0d_{3/2}}=(0.749 \pm 0.010 , -0.044 \pm 0.003)$ MeV, given with respect to the core $^{24}$O. Using the diffuseness and reduced radius from Ref. \cite{2016Hagino}, $a=0.72$ fm, $r_0=1.25$ fm, and the above experimental $\varepsilon_{1s_{1/2}}$ and $\varepsilon_{0d_{3/2}}$ states, we found (using $\chi^2$ optimization), $V_0=44.1$ MeV and $V_{so}=22.84$ MeV fm, for the Woods-Saxon and spin-orbit strengths, respectively. The calculated complex energy are, $\varepsilon_{1s_{1/2}}=-4.087$ MeV, $\varepsilon_{0d_{3/2}}=(0.749,-0.0436)$ MeV, $\varepsilon_{1p_{3/2}}=(0.576,-0.812)$ MeV and $\varepsilon_{0f_{7/2}}=(2.427,-0.102)$ MeV. Both the mean-field strengths and energies are very similar to the ones in Ref. \cite{2016Hagino}.
 
The path $L^+$, Fig. \ref{fig.cont1}(b), is chosen the same for all partial waves in order to have all three unperturbed resonant energies, $2\varepsilon_{0d_{3/2}}$, $2\varepsilon_{1p_{3/2}}$, and $2\varepsilon_{0f_{7/2}}$ within the zone free of non-resonant continuum-continuum states. Since the ground state of $^{26}$O is a threshold state, it is convenient to take the vertex $a=0$, and the other parameters $c=2$ MeV, $b=6$ MeV, and $d=100$ MeV. For the complex energy scattering states we include the partial waves from zero to four, then there are nine contours. In order to set up the discretization of the contours we take a strength $G$ which gives a loosely bound state of a few keV. Even when the basis is fully complex, the energy of the bound state must be real, so we take the criterion that imaginary part of the calculated energy be $\lesssim 10^{-8}$ MeV. Each contour associated to the three resonances, $d_{3/2},\, f_{7/2},\, p_{3/2}$ were discretized with 200 mesh-points, while each one of the other partial waves were discretized by 160 mesh-points. Then, the single-particle basis is form by three resonances and 1560 discrete complex-energy scattering states.

\subsection{Tracking the $^{26}$O resonances} \label{sec.4}
The zero-order two-particle states, formed by the ordered sum of the above complex energies states, covers a big region of the energy plane with a rectangular region free of contour-contour states, like in Fig. \ref{fig.cont2}(b). For $G=0$ the pole states of the three-body system are located at 
$2\varepsilon_{0d_{3/2}}=(1.498 ; -0.087)$ MeV, 
$2\varepsilon_{1p_{3/2}}=(1.152; -1.624)$ MeV, and 
$2\varepsilon_{0f_{7/2}}=(4.854, -0.204)$ MeV. When the interaction is turn on, these poles will move and the expectation is that some of them become physically meaningful states. The ordered two-particle basis has thousands of states, of which only three are of interest. Here one can appreciate the advantage of the separable force because finding roots is easier than diagonalizing a huge complex matrix. Furthermore, the choice of a rectangular path makes the identification of the states of interest visually simple.

\begin{figure}[h!t]
  \includegraphics[angle=0,width=0.45\textwidth]{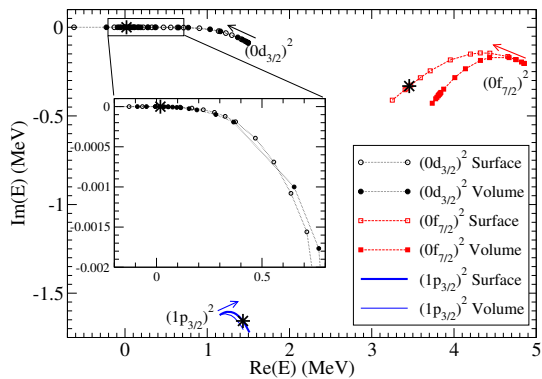}
\caption{Trajectory of the correlated poles as a function of the separable strength. The arrows indicate the direction in which $G$ increases with $0 \le G \le 7$ MeV. The large starts indicate the energies at the physical strength $G_{\rm{Exp}}=6.06$ MeV. The inset shows the trajectory for the $(0d_{3/2})^2$ state for Re($E$) $\lesssim 0.7$ MeV.}
\label{fig.moving}
\end{figure}

For the parameters of the residual interaction we adopt the same as the Woods-Saxon, namely $R=3.606$ fm and $a=0.72$ fm, for both, the surface and volume form factors $f(r)$. Figure \ref{fig.moving} shows the trajectory of the three correlated poles, i.e. the correlated two-body states whose main pole-pole contribution is one of the zeroth order configuration $(0d_{3/2})^2$, $(1p_{3/2})^2$ or $(0f_{7/2})^2$.
The evolution given by the two interactions are qualitatively the same. The inset shows details for the $(0d_{3/2})^2$ pole in the transition from resonance to the bound state.

Figure \ref{fig.moving} shows that the first resonance $(0d_{3/2})^2$ quickly moves to the two-particle continuum threshold, decreasing its real and imaginary parts, until it becomes a Borromean loosely bound state. Surface and volume pair interactions give the same result for this narrow resonance, as it can be seen in the inset. The trajectories of the poles $(0f_{7/2})^2$ and $(1p_{3/2})^2$ differ quantitatively but not qualitatively for the two interactions. Both poles slightly decrease their width when the interaction is switch on; then, after reaching a minimum (in absolute value), the width increases. In order to understand this behavior we show, in Tables \ref{table.wf-f} and \ref{table.wf-p}, the occupation numbers in the vicinity of the minimum. The configurations are separated as pole-pole (for example $(0d_{3/2})^2$), pole-scattering (for example $(0d_{3/2})(d_{3/2})$) and scattering-scattering $(d_{3/2})^2$. One observes that as the strength increases, the occupations of the pole-scattering and scattering-scattering increase in detriment of the  pole-pole configuration. As a final observation, the pole $(1p_{3/2})^2$ moves in the opposite direction with respect to the others poles. This intriguing behavior is a consequence of the combined effects of the Berggren metric, which replaces matrix elements $|M_{kl}|^2$ by $M_{kl}^2$ and the fact that $1p_{3/2}$ is a wide resonance. As a consequence the real part of $M_{kl}^2$ is negative and so, the effective interaction behaves as repulsive.

\begin{table*}[h!tb]
  \caption{\label{table.wf-f} Wave function occupation ($\%$) for the pole $(0f_{7/2})^2$ at the  minimum (in absolute value) of width and its neighborhood.}
\resizebox{\textwidth}{!}{%
\begin{tabular}{c|ccc|ccc|ccccccccc}
\hline \hline
  $E$ (MeV)         & $(0d_{3/2})^2$           & $(1p_{3/2})^2$          & $(0f_{7/2})^2$
  						   & $(0d_{3/2})(d_{3/2})$ & $(1p_{3/2})(p_{3/2})$ & $(0f_{7/2})(f_{7/2})$
  						   & $(s_{1/2})^2$            & $(p_{3/2})^2$             & $(p_{1/2})^2$
  						   & $(d_{5/2})^2$           & $(d_{3/2})^2$             & $(f_{7/2})^2$ 
  						   & $(f_{5/2})^2$            & $(g_{9/2})^2$             & $(g_{7/2})^2$  \\
\hline
  (4.654,-0.165) & (0.1,-0.1)         & (0.,0.)                          & (99.9,0.1)
                          & (0.,0.)                     & (0.,0.)                          & (0.,0.)
                          & (0.,0.)                     & (0.,0.)                         & (0.,0.)
                          & (0.,0.)                     & (0.,0.)                         & (0.,0.)
                           & (0.,0.)                    & (0.,0.)                         & (0.,0.) \\
  (4.366,-0.144) & (1.1,-0.2)       & (-0.2,0.1)           & (99.4,0.2) 
                          & (-0.1,-0.2)     & (-0.1,-0.1)                & (-0.3,0.1)
                          & (0.,0.)                     & (0.,0.)                         & (0.,0.1)
                          & (0.,0.)                     & (0.,0.)                         & (0.,0.)
                           & (0.1,0.)              & (0.1,0.)                   & (0.,0.) \\
  (4.079,-0.164) & (3.4,-0.2) & (-0.5,0.3)                   & (98.0,0.2) 
                          & (-0.3,-0.6)     & (-0.4,-0.1)           & (-0.6,0.3)
                          & (0.,0.)                     & (0.,-0.1)                  & (0.,0.2)
                          & (0.1,0.)                     & (0.,0.)                   & (0.,0.)
                           & (0.1,0.)              & (0.2,0.)                   & (0.,0.) \\
\hline \hline
\end{tabular}
} 
\end{table*} 
\begin{table*}[h!tb]
	\caption{\label{table.wf-p} Like Table \ref{table.wf-f} for pole $(1p_{3/2})^2$.}
\resizebox{\textwidth}{!}{%
\begin{tabular}{c|ccc|ccc|ccccccccc}
\hline \hline
  $E$ (MeV)         & $(0d_{3/2})^2$           & $(1p_{3/2})^2$          & $(0f_{7/2})^2$
  						   & $(0d_{3/2})(d_{3/2})$ & $(1p_{3/2})(p_{3/2})$ & $(0f_{7/2})(f_{7/2})$
  						   & $(s_{1/2})^2$            & $(p_{3/2})^2$             & $(p_{1/2})^2$
  						   & $(d_{5/2})^2$           & $(d_{3/2})^2$             & $(f_{7/2})^2$ 
  						   & $(f_{5/2})^2$            & $(g_{9/2})^2$             & $(g_{7/2})^2$  \\
\hline
  (1.174,-1.617) & (0.,0.1)                    & (100,-0.1)                          & (0.,0.)
                          & (0.,0.)                     & (0.,0.)                          & (0.,0.)
                          & (0.,0.)                     & (0.,0.)                         & (0.,0.)
                          & (0.,0.)                     & (0.,0.)                         & (0.,0.)
                           & (0.,0.)                    & (0.,0.)                         & (0.,0.) \\
  (1.263,-1.605) & (1.9,0.1)       & (98.4,-1.2)           & (-0.4,0.7) 
                          & (0.2,0.4)     & (0.1,0.1)                & (-0.1,-0.1)
                          & (0.,0.)                     & (0.,0.)                         & (-0.1,0.)                       
                          & (0.,0.)                     & (0.,0.)                         & (0.,0.)                         
                          & (0.,0.)              & (0.,0.)                   & (0.,0.) \\
  (1.339,-1.615) & (4.9,-1.5)      & (95.2,-1.5)            & (-0.7,2.1) 
                          & (0.7,0.8)        & (0.4,0.2)                & (-0.3,-0.2)
                          & (0.,0.)                     & (0.1,0.1)               & (-0.1,0.)                       
                          & (0.,0.)                     & (0.,0.)                           & (0.,0.)                        
                          & (-0.1,0.)             & (-0.1,0.)                   & (0.,0.) \\
\hline \hline
\end{tabular}
}
\end{table*}

\subsection{Physically meaningful resonances} \label{sec.exp}
In the previous section we studied the trajectories of the poles of the three-body Hamiltonian as a function of the strength. In this section we analyze each one of them for the physical strength, $G_{\rm{Exp}}=6.06$ MeV, that reproduces the experimental ground state energy $E=18$ keV \cite{2016Kondo}.

Table \ref{table.E} compares our calculated $0^+$ states for the surface interaction, with the theoretical models of Refs. \cite{2015Grigorenko,2016Hagino}. We can see a remarkable agreement with Ref. \cite{2016Hagino}. Our calculation finds the $0^+_2$ state close in energy to the one in Ref. \cite{2015Grigorenko}, but its width is so large that it prevents it to be a physical meaningful resonance. 

Experimentally, Ref. \cite{2013Kohley} founds the first excited state around $2$ MeV, while Ref. \cite{2013Caesar} founds it around $4.3$ MeV. Both experimental results give a width of approximately $1$ MeV, in good agreement with the imaginary part of the energy for the state $0^+_3$ in Table \ref{table.E}.

\begin{table}[h!tb]
\begin{ruledtabular}
	\caption{\label{table.E} Comparison of the calculated energies (MeV) of the $0^+$ states for $G_{\rm{Exp}}=6.06$ MeV with the models of Refs. \cite{2015Grigorenko,2016Hagino}.}
\begin{tabular}{c|ccc}
  state          &    This paper  & Ref.\cite{2016Hagino} & Ref.\cite{2015Grigorenko} \\
\hline
  $0^+_1 $  & $(0.0184,-0.164 \times 10^{-06})$ & $0.018$ & $0.01$   \\
  $0^+_2 $  & $(1.434,-1.657)$                             &  & $1.7$     \\
  $0^+_3 $  & $(3.460  ,-0.332)$                           & $(3.38,-0.366)$ & $2.6$     
\end{tabular}
\end{ruledtabular}
\end{table} 

The energy of the ground and first two excited states at the physical strength $G_{\rm{Exp}}$ are,
\begin{align*}
   E_{\rm{gs}} &= 0.01842-i\, 0.1635 \times 10^{-06} \; \rm{MeV}  \\
   E_{0^+_2} &=1.434- i\, 1.6568 \; \rm{MeV}  \\
   E_{0^+_3} &= 3.4560- i\, 0.3317 \; \rm{MeV} 
\end{align*}

The ground state wave function is,
\begin{align*}
 | ^{26} \rm{O} \rangle_{\rm{gs}} &= 
     (-0.862,0.060) | (0d_{3/2})^2 \rangle \\
     &+ (0.425,-0.033) | (0f_{7/2})^2 \rangle  \\
     &+ (0.305,-0.145) | (1p_{3/2})^2 \rangle \\
     &+(0.031,0.052) | (1p_{3/2})(c_1 p_{3/2}) \rangle \\
     &+ (0.028,0.054) | (1p_{3/2})(c_2 p_{3/2}) \rangle 
     + \cdots 
\end{align*}
We observe that the most important configurations for the collective state comes from the three pole-pole configurations.  Then, the next most important contributions comes from the $p_{3/2}$ shell, with one particle in the pole and the other in the complex energy scattering states at the energies $\varepsilon(c_1 p_{3/2})=-i0.907$ MeV, and $\varepsilon(c_2 p_{3/2})=-i0.845$ MeV.

The wave function of the first excited state has more single-particle character, with the main configuration $(1p_{3/2})^2$. As, for the ground state, the other pole-pole configurations are also relevant. The next most important configuration comes from the $d_{3/2}$ shell, with one neutron in the resonant continuum and the other in the non-resonant continuum at the energies $\varepsilon(c_1 d_{3/2})=0.767-i2$ MeV, and $\varepsilon(c_2 d_{3/2})=0.646-i2$ MeV.
\begin{align*}
    | ^{26} \rm{O} \rangle_{0^+_2} &= 
     (-0.306,0.118) | (0d_{3/2})^2 \rangle \\
     & + (0.161,0.156) | (0f_{7/2})^2 \rangle \\
     & + (-0.951,-0.004) | (1p_{3/2})^2 \rangle \\
     & + (0.130,-0.018) | (0d_{3/2})(c_1 d_{3/2}) \rangle \\
     & + (0.126,0.021) | (0d_{3/2})(c_2 d_{3/2}) \rangle 
     + \cdots 
\end{align*}

The second excited state, is also non collective, having as main configuration the two neutrons in the $0f_{7/2}$ resonant continuum.
\begin{align*}
  | ^{26} \rm{O} \rangle_{0^+_3} &= 
     (0.409,0.072) | (0d_{3/2})^2 \rangle \\
     &+ (0.945,-0.030) | (0f_{7/2})^2 \rangle \\
     &+ (0.027,0.160) | (1p_{3/2})^2 \rangle \\
     &+ (0.028,-0.049) | (0d_{3/2})(c_1 d_{3/2}) \rangle \\
     &+ (0.022,-0.052) | (0d_{3/2})(c_2 d_{3/2}) \rangle
     + \cdots 
\end{align*}
The most important amplitude from the non resonant contribution comes from the $d_{3/2}$ shell, with one particle in the resonant state and the other in the scattering state. The first two configurations are the ones at the energies $\varepsilon(c_1 d_{3/2})=2.535-i2$ MeV, and $\varepsilon(c_2 d_{3/2})=2.720 -i2$ MeV. The energies of the complex energy scattering states most important are the ones which, when summed to the pole energy, give a figure close to the correlated energy. This is a characteristic of the separable interaction, which can be seen from Eq. (\ref{eq.x}); for example, 
$\varepsilon(0d_{3/2})+\varepsilon(c_1 d_{3/2})=(0.749-i\, 0.0436) + (2.535-i2.) =3.284 - i\, 2.044$ MeV, while the correlated energy is $3.456- i\, 0.332$ MeV; the difference with the imaginary part of the correlated energy is because the remainder contour also largely contribute to the width. 

Tables \ref{table.v1} and \ref{table.v3} compare our calculated occupation probabilities of the two meaningful resonances with those of Refs. \cite{2015Grigorenko,2016Hagino}. The symbol $\sum$ indicates that the summation of the pole-pole, pole-scattering and scattering-scattering has been performed. 

For the ground state (Table \ref{table.v1}), the comparison with Ref. \cite{2015Grigorenko}, shows that the occupation of the $s$ state is of the same order of magnitude, but they differ appreciable for the $d_{5/2}$ configuration; while comparison with Ref. \cite{2016Hagino} shows that the main contributions come from the configurations $d_{3/2}$, $f_{7/2}$ and $p_{3/2}$, with an excellent agreement for the occupation of the $f_{7/2}$ shell. Finally, our model predicts bigger occupation for the configuration $d_{3/2}$ in detriment of the $p_{3/2}$ one, giving a somewhat less collective ground state.

\begin{table*}[h!tb]
\begin{ruledtabular}
	\caption{\label{table.v1} Occupation probabilities in $\%$ for the state $0^+_1$ at the energy $E_1=0.01842-i\, 0.1635 \times 10^{-6}$ MeV.}
\begin{tabular}{c|ccccccccc}
  Model       & $\sum (s_{1/2})^2$ & $\sum (p_{1/2})^2$ & $\sum (p_{3/2})^2$
  				   & $\sum (d_{3/2})^2$ & $\sum (d_{5/2})^2$ & $\sum (f_{5/2})^2$ 
  			       & $\sum (f_{7/2})^2$  & $\sum (g_{7/2})^2$ & $\sum (g_{9/2})^2$  \\
\hline  						   
  This paper  & (0.2 ,0.)  &  (0.5 ,0.) & (4.2 ,0.) 
                   & (73.7 ,0.)  & (0.3 ,0.) & (0.7 ,0.) 
                   & (19.0 ,0.)  & (0.3 ,0.)  & (1.1 ,0.) \\
  Ref.\cite{2015Grigorenko}
                  &  0.67                             &                               &       
                  &  79                         &     19                          &      
                  &                           &                               & \\
  Ref.\cite{2016Hagino}
                  &                                  &                               &       10.5
                  &  66.1                         &                               &      
                  &   18.3                        &                               & 
\end{tabular}
\end{ruledtabular}
\end{table*} 

The comparison of the $0^+_3$ state (Table \ref{table.v3}) with Refs. \cite{2015Grigorenko} and \cite{2016Hagino} shows almost the same features as for the ground state, with two differences. First, the great occupation of the $s$ shell in Ref. \cite{2015Grigorenko} and second, the smaller collectivity of our wave function with respect to that of Ref. \cite{2016Hagino}. This last feature may be explained by the structure of the analytic expression of the wave function amplitude Eq. (\ref {eq.x}), which favors the collectivity of the ground state, while it inhibits the collective character of excited states. 

\begin{table*}[h!tb]
\begin{ruledtabular}
	\caption{\label{table.v3} Like Table \ref{table.v1} for the state $0^+_3$ at the energy $E_3=3.4560- i\, 0.3317$ MeV.}
\begin{tabular}{c|ccccccccc}
  Model       & $\sum (s_{1/2})^2$ & $\sum (p_{1/2})^2$ & $\sum (p_{3/2})^2$
  				   & $\sum (d_{3/2})^2$ & $\sum (d_{5/2})^2$ & $\sum (f_{5/2})^2$ 
  			       & $\sum (f_{7/2})^2$  & $\sum (g_{7/2})^2$ & $\sum (g_{9/2})^2$  \\
\hline  						   
  This paper  & (0., 0.1)  &  (0.1, 0.6) & (-3.7, -0.1) 
                   & (15.4, 3.7)  & (0.2, 0.1) & (0.4, 0.1) 
                   & (86.9, -4.7)  & (0.1, 0.)  & (0.6, 0.2) \\
  Ref.\cite{2015Grigorenko}
                  &  3.8                             &                               &       
                  &  86                         &     6.1                         &      
                  &                           &                               & \\
  Ref.\cite{2016Hagino}
                  &                                  &                               &       10.4
                  &  24.9                         &                               &      
                  &   62.1                        &                               & 
\end{tabular}
\end{ruledtabular}
\end{table*} 

Finally, from the imaginary part of the calculated ground state energy we get a half live 
$T_{1/2}=\frac{\hbar \ln2}{-2 {\rm Im} (E(O^+_1))}= 8.766 \times 10^{-3}$ ps, which is three orders of magnitude smaller than the experimental one $4.5 \pm 3$ ps \cite{2013Kohley}. Since from Table \ref{table.v1} we can argue that $l_{\rm{max}}=4$ is large enough, an improvement in the model may require a better treatment of the N-N residual interaction, specifically the continuum coupling.

\section{Conclusions}\label{sec.conclusion}
The structure of the $0^+$ states of the $^{26}$O nucleus have been studied in the complex energy plane with a separable interaction. Continuum partial waves up to $l=4$ have been used and their contribution to the ground and excited states were examined as a function of the pair interaction strength $G$. The trajectory of the ground state shows that a little stronger residual interaction may produce a loosely bound Borromean nucleus. This subtle balance between the pair interaction and the continuum makes this nuclei very hard to quantitatively assess. One of the resonances was discarded because its width does not decrease significantly with the interaction and the real part of its energy does not follow the usual behavior of the resonances. At the physical strength $G_{\rm{Exp}}=6.06$ MeV, our wave function amplitudes are in agreement with that of Ref. \cite{2016Hagino}, while the physically meaningful excited state has the following resonant parameters, $3.46-i\, \frac{0.66}{2}$ MeV, similar to the result of Ref. \cite{2016Hagino}. The  calculated half-live of the ground state, obtained from the imaginary part of its complex energy, is three-orders of magnitude smaller than the experimental one. We interpret this as an indication that some correlations are still missing and suggest that a more realistic or phenomenological interaction adjusted for the continuum \cite{2017Yaganathen} might be required.

\begin{acknowledgments}
This paper has been supported by the National Council of Research PIP-625 and the University of Rosario ING588, Argentina.
\end{acknowledgments}


%

\end{document}